\documentclass[twocolumn,prl]{revtex4}
\usepackage{amsfonts,amssymb,amsmath}
\usepackage{eurosym}
\usepackage{graphicx}

\begin{document}

\title{Comment on ``The renormalized superperturbation theory (rSPT) approach
to the Anderson model in and out of equilibrium''}
\author{A. A. Aligia}
\affiliation{Centro At\'{o}mico Bariloche and Instituto Balseiro, Comisi\'{o}n Nacional
de Energ\'{\i}a At\'{o}mica, CONICET, 8400 Bariloche, Argentina}
\date{\today}
\maketitle

Here I present briefly some facts about 
nonequilibrium renormalized perturbation theory (RPT), correcting recent misleading statements in 
Ref. \onlinecite{jpcs}, and discuss some results of this work using rSPT at equilibrium.

In RPT, the {\bf bare renormalized} local spectral density from which renormalized retarded $\tilde{\Sigma}^{r}(\omega)$ 
and lesser $\tilde{\Sigma}^{<}(\omega )$ self energies are calculated is given by \cite{hew,sca} 

\begin{equation}
\tilde{\rho }_{0}(\omega )=\frac{\tilde{\Delta }/\pi }
{(\omega -\tilde{\varepsilon }_{d})^{2}+\tilde{\Delta }^{2}},  \label{rhor}
\end{equation}
where $\tilde{\varepsilon}_{d}$ is the renormalized d level, which using Friedel sum rule 
(FSR) for zero temperature ($T$) and voltage ($V$) can be related to the local occupancy $n_d$ \cite{hew,sca}: 

\begin{equation}
\tilde{\varepsilon}_{d}=\tilde{\Delta}\cot (\pi n_{d}/2),
\label{edeff}
\end{equation}
where here zero applied magnetic field $B$ is assumed.

In Ref. \onlinecite{jpcs}, Mu\~{n}oz et al. review some results of
equilibrium rSPT. In the appendix they discuss a Ward identity (WI) in general. In
particular, they argue that in Ref. \onlinecite{sca}, $\tilde{\rho }_{0}(\omega )$ contains
a voltage dependence that invalidates the WI. This statement is
unfounded. In fact Eq. (\ref{edeff}) [(11) in Ref. \onlinecite{sca}] 
was derived using FSR \cite{sca}, which is only valid for $T=V=0$. 
Therefore to assume that $n_d$ is calculated
selfconsistently [see Eq. (52) of Ref. \onlinecite{jpcs}] in Ref. \onlinecite{sca} to modify 
$\tilde{\rho }_{0}$ is preposterous. Moreover in Ref. \onlinecite{sca}, it is stated that
``{\it The parameters of the original model are renormalized and their values can
be calculated exactly from Bethe ansatz results [44--48], or accurately
using NRG [49--53]}'' (both techniques are only valid at $V=0$) and ``{\it an
accurate knowledge of thermodynamic quantities from the Bethe ansatz or NRG
permits a precise determination of $z=\rho (0)/\tilde{\rho }_{0}(0)$,
and the renormalized interaction through the Wilson ratio 
$R=1+\tilde{U}\tilde{\rho }_{0}(0)$.}''  The constant $\tilde{\rho }_{0}(0)$  
enters the explicit
expression of $\tilde{\Sigma}^{<}(\omega )$ 
given by Eq. (20)  of Ref. \onlinecite{sca} (E20R3). E20R3   
was shown to satisfy the WI in
Ref. \onlinecite{rr}. The results were generalized to finite temperature in Ref. \onlinecite{ng}, where 
also the  nonequilibrium heat current of a
nanodevice was discussed using renormalized parameters $\tilde{\Delta}$, $\tilde{\varepsilon}_{d}$ and 
$\tilde{U}$ obtained from NRG following the method of
Hewson \textit{et al}. \cite{hom} as described in Ref. \onlinecite{cb}, where also
$\tilde{\Delta}$, $\tilde{\varepsilon}_{d}$ and 
$\tilde{U}$ for several values of the
original parameters were tabulated. In all these works of course 
$\tilde{\varepsilon}_{d}$ is constant independent of voltage and temperature.

Previous results of some of the authors \cite{prl}
claimed that $\tilde{\Sigma}^{<}(\omega )= 2i\tilde{f}(\omega)$Im$(\tilde{\Sigma}^{r}(\omega))$,
where $\tilde{f}(\omega)$ is an average of the Fermi function at the two leads.
This expression is incorrect and leads to spurious jumps in $\tilde{\Sigma}^{<}(\omega )$
at $T=0$ \cite{com}. 
In Ref. \onlinecite{jpcs} this result is corrected. 
In fact {\bf Eq. (41) for $\tilde{\Sigma}^{<}(\omega )$ of Ref. \onlinecite{jpcs} 
is identical to Eq. (16) of Ref. \onlinecite{sca} and when evaluated to order $V^2$ 
at $T=0$ leads to the correct result, E20R3}. 
For finite small $T$, $\tilde{\Sigma}^{<}(\omega )$ and Im$(\tilde{\Sigma}^{r}(\omega))$ are
evaluated in Ref. \onlinecite{ng}.

The repeated attempts of Mu\~{n}oz and Kirchner to undermine Ref. \onlinecite{sca} have been inconsistent 
over the time. 

First, in Ref. \onlinecite{prl} they stated ``{\it A problem with this approach is that it fails to recover
$p-h$ symmetry at $\tilde{u}=1$ and gives a linear in $T$ term in the
spectral density away from half filling $n=1$, in contradiction to certain Ward identities}.''  
However, Eq. (30) of Ref. \onlinecite{sca} precisely shows that previous rigorous results for 
$n_d=1$ \cite{sela} are recovered, including the particle-hole ($p-h$) symmetric case. 
Concerning the $T$ dependence, Ref. \onlinecite{sca} is dedicated to $T=0$.
In the small section 3.5 a brief comment is given on the effect of the Hartree term
on the {\bf dressed unrenormalized} spectral density $\rho (\omega)$ (which {\bf does not} enter the WI's)
in self-consistent ordinary (not renormalized) perturbation schemes \cite{sca} (b),
but this has not been used and in any case, terms linear in $T$ are absent for $V=0$. 
The extension to finite $T$ is done in Ref. \onlinecite{ng}.

Second, in Ref. \onlinecite{reply} they ``proved'' that $\tilde{\Sigma}^{<}(\omega )$ for $T=0$ given by 
E20R3 does not satisfy the WI
based on a (wrong) expansion of $\tilde{\Sigma}^{<}(\omega )$ around $\omega=V=0$ 
(no mention of the dependence of $\tilde{\varepsilon}_{d}$ on $V$ alleged in Ref. \onlinecite{jpcs} 
was made at this stage). They state the violation of the WI in Refs. \onlinecite{sca,com} 
``{\it is most clearly seen by noticing the linear-in-$\omega$ and linear-in-$V$ terms in 
} E20R3''.
This point is clarified in Ref. \onlinecite{rr} and Ref. 29 of Ref. \onlinecite{ng}.
The derivatives involved in the WI were calculated explicitly for $T=0$ 
(Ref. \onlinecite{rr}) and arbitrary $T$ (Ref. \onlinecite{ng}),
showing that the WI's are fulfilled.

Third, as explained above, in Ref. \onlinecite{jpcs} they now take the same expression
[Eq. (16) of Ref. \onlinecite{sca}]
that leads to E20R3, but invent that $\tilde{\rho }_{0}(\omega )$
contains a dependence on $V$.

As shown above, the nonequilibrium RPT scheme, which 
in the most complete form is given in Ref. \onlinecite{ng}, is correct and satisfies the 
WI's. However, it has important limitations. One of them is that it is restricted to 
$eV \ll k_B T_K$. For this reason, alternative approaches are usually preferred, 
like the non-crossing approximation (see e.g. Refs. \onlinecite{win,serge,capac,nca,roura}), which
reproduces well the scaling relations with temperature $T$ and $V$ in the Kondo regime 
\cite{roura}.

Ref. \onlinecite{jpcs} presents also equilibrium results using rSPT for the first term
in the expansion of the conductance as a function of $(T/T_0^s)^2$ ($c^\prime_T$) and also
a function of $(g \mu_B B/k_B T_0^s)^2$ ($c^\prime_B$), where 
the Kondo scale is defined as $T_0^s=(g \mu_B)^2/(4 \chi(0))$, being $\chi(0)$ the magnetic susceptibility
for the symmetric Anderson model ${\varepsilon}_{d}=-U/2$, for which $n_d=1$.  
I use the notation of a previous paper of the authors on the subject
[Eqs. (20) and (21) of Ref. \onlinecite{cbm}; see also Ref. \onlinecite{cb} for some corrections to this work]. 
One realizes that including a sum of ladder diagrams the results presented improve considerably when compared to NRG 
results. However, as I argue below, the results are still somewhat disappointing.

One point to be noted is that out of the symmetric point ${\varepsilon}_{d}=-U/2$, it would be more natural
to use the energy scale $T_0=(g \mu_B)^2/(4 \chi)$ with $\chi$ calculated for the actual value 
of ${\varepsilon}_{d}$ to define the expansion coefficients. This leads to $c_T$ and $c_B$ as defined in 
Ref. \onlinecite{cbm}.
The symmetric point, and therefore $T_0^s$ might not be experimentally accessible. This is the case of
some molecular system in which $U$ is very large \cite{serge}. Moreover, since $\chi$ decreases ($T_0$
increases) fast when moving to the intermediate valence region ${\varepsilon}_{d} \sim 0$, the $c^\prime$ 
are considerably smaller (by a factor $(T_0^s/T_0)^{2}$) than the $c$. As a consequence, while
the $c$ have an increasing downward curvature as ${\varepsilon}_{d}$ increases from $-U/2$ to $0$ \cite{cb},
the $c^\prime$ have an inflexion point \cite{jpcs} and become in general much smaller for 
${\varepsilon}_{d} \sim 0$. Since 
at the other end of the plotted values,  $\varepsilon_{d}=-U/2$ 
the coefficients are fixed by Fermi liquid properties, 
the end points of $c^\prime$ are rather fixed and
it is hard to see deviations from any two curves of $c^\prime$,  while they are more evident if
$c$ is represented. In Ref. \onlinecite{jpcs} the authors plot with the names $c_T$ and $c_B$ what they 
had called $c^\prime_T$ and $c^\prime_B$  in Ref. \onlinecite{cbm}.

\begin{figure}[tbp]
\includegraphics[width=7. cm]{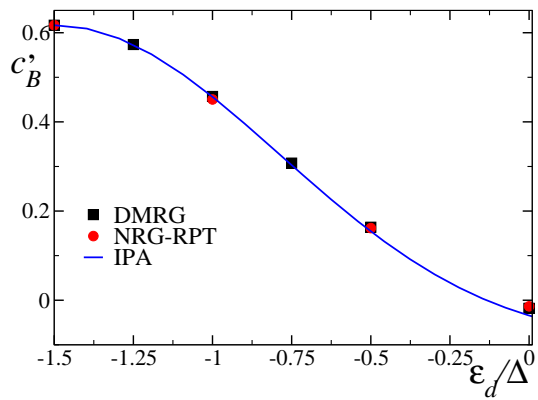}
\caption{Coefficient $c^\prime_B$ of $G/G_0=1-c^\prime_B(g \mu_B B/k_B T_0^s)^2$ vs 
$\varepsilon_{d}$ for $U=3 \Delta$ obtained by different methods.}
\label{cb}
\end{figure}

In Fig. \ref{cb}, I show the results for $c_B$ represented in Fig. 1 of Ref. \onlinecite{cb} 
rescaled by the factor $(T_0^s/T_0)^{2}$ to represent $c^\prime_B$. The results of a very simple 
interpolative perturbative approach (IPA) \cite{kk} seem excellent and better than the rSPT results shown in Fig. 2 
of Ref. \onlinecite{jpcs}, particularly for $\varepsilon_{d} <  -0.5 \Delta$. 
However, in Fig. 1 of Ref. \onlinecite{cb}
one can see that the IPA results are systematically lower than the more precise obtained using 
DMRG or combining NRG and RPT. The origin of this difference is twofold, 
errors in evaluating $T_0$ within IPA and the 
factor $(T_0^s/T_0)^{-2}$. The difference between IPA and DMRG results for $c_B$ increases as $\varepsilon_{d}$ 
moves away of the symmetric point and reaches 12\% of the maximum value of $c_B$ for $\varepsilon_{d}=0$. 
This difference is reduced to 2.6\% when $c^\prime_B$ is plotted. 

While the IPA might be considered acceptable for $U= 3 \Delta$ and improves considerably as $U$ 
is lowered \cite{cb}, the
main problem is that 
the IPA \cite{cb} (and it seems to be also the case rSPT \cite{jpcs}) rapidly deteriorates as $U$ increases. In the 
Kondo limit $-\varepsilon_{d}, \varepsilon_{d}+U \gg \Delta$, one knows that the spectral density displays 
two charge-transfer peaks for $\omega \sim \varepsilon_{d}$ and $\omega \sim \varepsilon_{d}+U$ of total width 
$4 \Delta$ and a Kondo peak at the Fermi level of width of the order of $2\tilde{\Delta}$ \cite{capac}.
For $U = 3 \Delta$ (the largest value of $U$ considered in Ref. \onlinecite{jpcs}), these peaks cannot 
be separated. More accurate methods seem necessary to treat the Kondo case \cite{cb}.

\end{document}